\documentclass[rsi,superscriptaddress,twocolumn,amsmath,amssymb,reprent]{revtex4-1}

\usepackage{bm}
\usepackage{graphicx}%
\usepackage[colorlinks]{hyperref}
\usepackage{hyperref}
\hypersetup{
    colorlinks=true,
    linkcolor=blue,
    filecolor=magenta,      
    urlcolor=cyan,
    citecolor=red,
}

\usepackage[version=3]{mhchem} 

\usepackage{amsmath}
\usepackage{siunitx}

\begin{document}

\title{Thermally Driven Approach To Fill Sub-10-nm Pipettes with Batch Production}






\author{Linhao Sun}
\altaffiliation{Contributed equally to this work}
\affiliation{Nano Life Science Institute (WPI-NanoLSI), Kanazawa University, Kakuma-machi, Kanazawa 920-1192, Japan}

\author{Kazuki Shigyou}
\altaffiliation{Contributed equally to this work}
\affiliation{Nano Life Science Institute (WPI-NanoLSI), Kanazawa University, Kakuma-machi, Kanazawa 920-1192, Japan}

\author{Toshio Ando}
\email[]{tando@staff.kanazawa-u.ac.jp}
\affiliation{Nano Life Science Institute (WPI-NanoLSI), Kanazawa University, Kakuma-machi, Kanazawa 920-1192, Japan}

\author{Shinji Watanabe}
\email[]{wshinji@se.kanazawa-u.ac.jp}
\affiliation{Nano Life Science Institute (WPI-NanoLSI), Kanazawa University, Kakuma-machi, Kanazawa 920-1192, Japan}

\begin{abstract}


Typically, utilization of small nanopipettes results in either high sensitivity or spatial resolution in modern nanoscience and nanotechnology.
However, filling a nanopipette with a sub-10-nm pore diameter remains a significant challenge.
Here, we introduce a thermally driven approach to filling sub-10-nm pipettes with batch production, regardless of their shape.
A temperature gradient is applied to transport water vapor from the backside of nanopipettes to the tip region until bubbles are completely removed from this region.
The electrical contact and pore size for filling nanopipettes are confirmed by current-voltage and transmission electron microscopy (TEM) measurements, respectively.
In addition, we quantitatively compare the pore size between the TEM characterization and estimation on the basis of pore radius and conductance.
The validity of this method provides a foundation for highly sensitive detection of single molecules and high spatial resolution imaging of nanostructures.

\end{abstract}

\maketitle 
\newcommand*\mycommand[1]{\texttt{\emph{#1}}}

\section{Introduction}

Nanopipettes have been widely utilized in modern nanoscience and nanotechnology applications, such as molecular sensing~\cite{steinbock2013dna,steinbock2010detecting,gong2013label,zhang2009silica,ivanov2015demand,gibb2014single}, chemical delivery~\cite{bruckbauer2002writing,babakinejad2013local,page2017quantitative,takahashi2011multifunctional}, and scanning probe microscopy~\cite{traversi2013detecting,venkatesan2011nanopore,sze2017single,wanunu2010electrostatic,storm2003fabrication,li2001ion,siwy2003electro}.
In these applications, a nanometer-scale channel that forms near the tip of the nanopipette must be filled with a solution.
Such a solution-filled channel can facilitate the transport, counting, and detection of nanometer-scale objects passing through the nanopipette pore, typically undertaken using electrical methods.
The size of the nanopipette pore is adjusted on the basis of the applications; in particular, scanning probe microscopy techniques using nanopipettes require a smaller pore size as smaller pore sizes correspond to improved spatial resolution~\cite{rheinlaender2015lateral,rheinlaender2017accurate}.
Scanning ion conductance microscopy~\cite{hansma1989scanning} (SICM) through the use of nanopipettes has received significant use in the field of scanning probe microscopy owing to its huge potential for mapping surface topography~\cite{zhou2018nanoscale,shevchuk2006imaging,novak2009nanoscale,rheinlaender2013mapping}, mechanical properties~\cite{zhou2018nanoscale,shevchuk2006imaging,novak2009nanoscale,rheinlaender2013mapping}, and surface charge distribution~\cite{mckelvey2014surface,mckelvey2014bias,page2016fast,perry2015simultaneous,perry2016surface,klausen2016mapping,fuhs2018direct} of biological materials with nanometer-scale resolution.
In SICM, an ion current passing through the nanopipette pore generated by electrolyte-filled nanopipettes is used as a signal to sense the interface position of measured samples. 
However, using nanopipettes with a 10-nm pipet pore is difficult in practice owing to the strong capillary force that prevents the filling of the nanopipette pore with the solution.

Several attempts have been reported to solve this problem.
One method applies high pressure from the backside of a nanopipette to force the electrolyte solution toward the tip region.
However, this method is not suitable for a small nanopipette~\cite{salanccon2018filling}.
Another method utilizes the nanopipettes with a glass filament inside.
%
Although the electrolyte solution can easily fill the tip region of the nanopipette, these nanopipettes are easily broken by the application of undesirable forces that are frequently and accidentally generated at the nanopipette tip during experimental operations.
Very recently, the Tinland group reported a method to fill a nanopipette with a pore diameter of less than \SI{50}{nm} without a filament inside, performing a microdistillation procedure~\cite{salanccon2018filling}.
However, their microdistillation method is very time consuming as only one nanopipette can be filled in any given operation, significantly limiting the application of this method.
In addition, there was no direct evidence of the application of this method for electrolyte-filled nanopipettes with pore diameters less than \SI{10}{nm}.
The methods that they used are an indirect estimation of the pore diameter, i.e., scanning electron microscope (SEM) and electrical measurements of nanopipettes, which was clearly pointed out by Perry et al~\cite{perry2016characterization}.
Therefore, a new approach to fill sub-10-nm pipettes is greatly desired for a wide range applications in nanoscience and nanotechnology.

In this study, we proposed a simple and efficient method for filling nanopipettes with batch production, including those with pore diameters of less than \SI{10}{nm}.
A temperature gradient is applied to transport water vapor from the backside of the nanopipette to the tip region until bubbles are completely removed from the nanopipettes.
Electrical measurements ensure that our method is capable of filling all nanopipettes, regardless of the geometry.
The combination of electrical measurements and a transmission electron microscopy (TEM) image of the nanopipette provided direct evidence of the validity of our method for sub-10-nm nanopipettes.


\begin{figure*}[!t]
\includegraphics[pagebox=artbox]{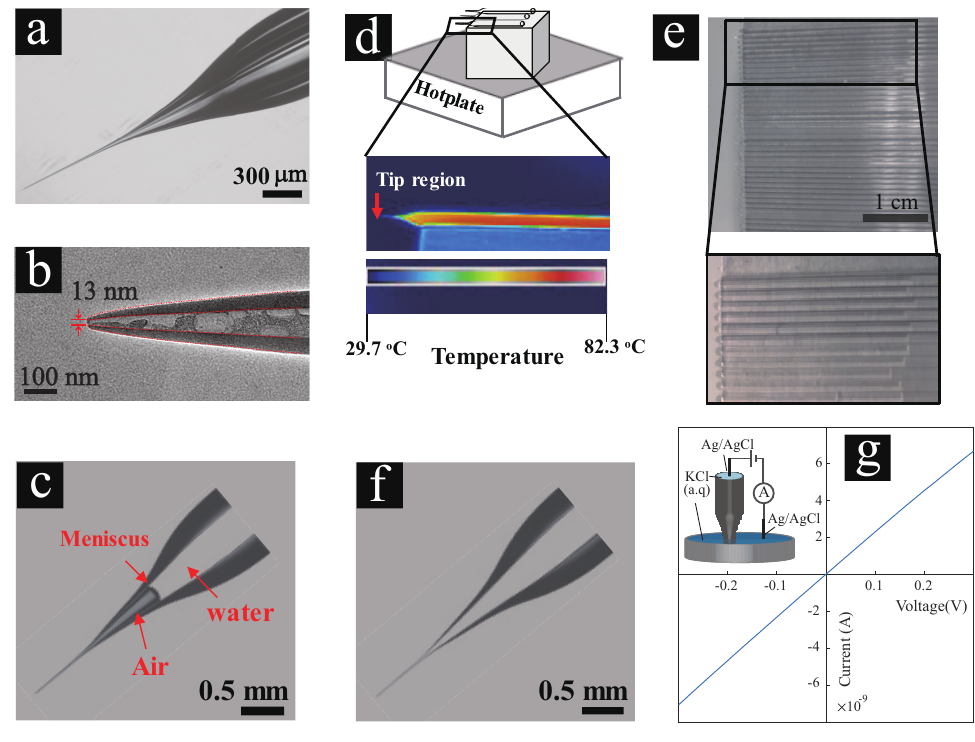}
\caption{\label{FIG1}
Experimental setup of thermally-driven (TD) method and the characterization of nanopipettes.
We used nanopipettes with a pore diameter of approximately \SI{10}{nm} to apply the TD method.
The geometry of the nanopipette used was characterized by the OM and TEM images.
The $I$-$V$ measurements were carried to ensure complete filling of solutions inside the nanopipettes.
(\textbf{a} and \textbf{b}) Example of the OM and TEM images of the nanopipettes used.
(\textbf{c}) OM image of a nanopipette after injecting DI water.
The meniscus forms at the water/air interface.
(\textbf{d}) Schematic of the TD method.
(\textbf{e}) Several tens of nanopipettes on metal-supported surface.
After being placed on the metal-supported surface, the nanopipettes were subjected to the temperature gradient generated by the hot plate.
The image of the temperature gradient was measured when the hotplate was kept at a temperature of $\sim$\SI{80}{\degreeCelsius}.
(\textbf{f}) OM image of the nanopipette after complete filling by the TD method.
The meniscus formed in panel \textbf{c} disappeared.
(\textbf{g}) Schematic of $I$-$V$ measurement of nanopipettes.
}
\end{figure*}

\section{Experimental Section}
\hypertarget{ES}{}
\subsubsection{Solutions and Materials}

Aqueous electrolyte solutions (\SI{2}{M} KCl, \SI{50}{mM} FeCl$_3$) were prepared with deionized water (DI H$_2$O; resistivity, ca. \SI{18}{\mega \Omega.cm} at \SI{25}{\degreeCelsius}; Millipore Corp.).
A non-metallic syringe needle (MF34G-5, World Precision Instruments) is used to fill glass pipettes with DI water or KCl solution.

\subsubsection{Nanopipette Fabrication}

In our experiments, all the pipettes were fabricated using a CO$_2$-laser puller (P-2000, Sutter Instrument).
The quartz capillaries utilized have an outer diameter o.d. = \SI{1.0}{mm} and inner diameter i.d. = \SI{0.3}{mm}.
Before the pulling process, all the capillaries were treated by O$_2$ plasma for \SI{10}{min} to remove some contamination inside of the nanopipettes.
The pulling parameters for the fabrication of sub-10-nm pipettes are shown in \href{https://pubs.acs.org/doi/suppl/10.1021/acs.analchem.9b03848/suppl_file/ac9b03848_si_001.pdf}{\textcolor{blue}{Supporting Information, Table S1}}. 

\subsubsection{$I$-$V$ Measurement}

The electrical properties of electrolyte-filled nanopipettes were characterized in a system fabricated in-house.
A current amplifier (DLPCA-200, FEMTO) was used to amply the ion current signal.
A function generator (WF-1973, NF Corp.) was used to supply a bias voltage between Ag/AgCl electrodes.
The Ag/AgCl electrode was made by immersing a silver wire into FeCl$_3$ solution for \SI{1}{min}.
An oscilloscope (DSOX1102G, Keysight Technologies) was utilized to record the input and output voltages.

The $I$-$V$ measurements were performed as follows.
A nanopipette was completely filled with DI by our proposed method.
The DI water inside the pipet was exchanged by a \SI{2}{M} KCl solution and kept for over \SI{10}{min}.
The nanopipette was immersed in a bath solution with a concentration of \SI{2}{M} KCl  and  kept  for \SI{10}{min}, inducing the formation of a homogeneous concentration in the nanopipette owing to the diffusion of ions.
Then, $I$-$V$ measurements were carried out with a bias voltage in the range from $-0.3$ to \SI{0.3}{V}.
The electrical resistance was calculated by linear fitting to the measured $I$-$V$ curve with a bias voltage in the range from $-50$ to \SI{50}{mV}.
The use of \SI{2}{M} KCl solution in the $I$-$V$ measurements helps to suppress the ion current rectification.
In addition, the small bias range to analyze the electrical resistances is useful to apply the linear fitting to the $I$-$V$ curves.    

\subsubsection{Optical Microscope Measurement}

All optical microscope (OM) images in this study were postprocessed by a painting software; the process is described in detail in the \href{https://pubs.acs.org/doi/suppl/10.1021/acs.analchem.9b03848/suppl_file/ac9b03848_si_001.pdf}{\textcolor{blue}{Supporting Information, Figure S1}}.

\subsubsection{TEM Measurement}
\label{TEM}

The TEM measurements were performed as follows:
After the $I$-$V$ measurement of the sub-10-nm pipet, electrolytes in the nanopipette were exchanged with the DI water.
The nanopipettes were stored in DI water for a day, and the nanopipette tips were cut and aligned with the copper TEM grid using the micromanipulator (Micro Support Co., Ltd.).
The aligned tips on the TEM grid were fixed by dropping ethanol.
The pore diameter of the nanopipettes was measured by TEM (JEM-2100Plus, JEOL Ltd.) with an accelerating voltage of \SI{200}{kV}.
All captured pore diameters were analyzed using Fiji (open source software ImageJ).


\subsubsection{Temperature Characterization}

An infrared thermometer (Fluke VT04A, Fluke Corp.) was used to measure the spatial profile of the temperature in the vicinity of the nanopipette.

\section{Results and Discussion}

\begin{figure}[!t]
    \includegraphics{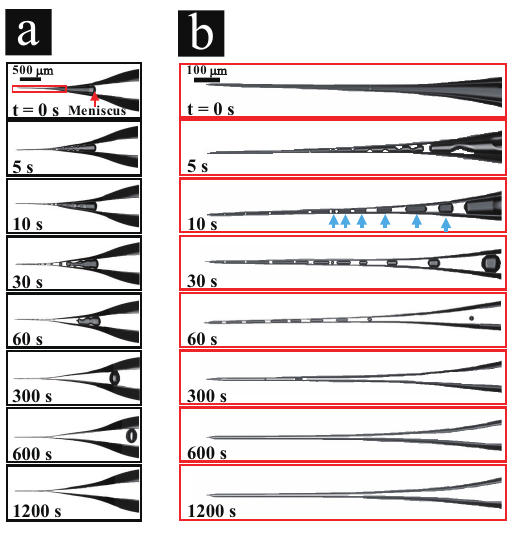}
    \caption{\label{FIG2}
    Time-lapsed images of the filling process of a nanopipette in the presence of the temperature gradient.
    The temperature of the hotplate was kept at $\sim$\SI{80}{\degreeCelsius} to generate the temperature gradient around the nanopipette tip.
    (\textbf{a} and \textbf{b}) Global and magnified images, respectively. The magnified area is indicated as the red square in panel \textbf{a}.
    }
\end{figure}

\textcolor{blue}{Figure \ref{FIG1}}a shows the optical microscope (OM) image of the fabricated nanopipette with a conical shape.
This nanopipette has a pore diameter of \SI{13}{nm} determined by the TEM image (\textcolor{blue}{Figure \ref{FIG1}}b).
When DI water is injected into the inside of the nanopipette, a meniscus forms at the water/air interface owing to the capillary force (\textcolor{blue}{Figure \ref{FIG1}}c).
The existence of an air domain prevents the filling of solutions in the nanopipette tip.
To remove the air domain, we propose ``thermal-driven (TD) method''.
Several prepared nanopipettes were placed on a metal-stage-supported hotplate, such that the sharp tip region of the nanopipettes protruded from the edge of the metal stage.
The protruding length of the nanopipettes was tuned in a range from \SI{2}{mm} to several mm, depending on the tip shapes of the nanopipettes.
In this situation, the temperature gradually decreased toward the pipette tip (\textcolor{blue}{Figure \ref{FIG1}}d), forming a temperature gradient near the tip region of the nanopipettes.
After maintain the temperature gradient for a while, the air domain can be removed from the nanopipette (\textcolor{blue}{Figure \ref{FIG1}}f).
The evidence of the complete removal of the air domain was obtained by  current-voltage ($I$-$V$) measurements (\textcolor{blue}{Figure \ref{FIG1}}g).
Note that the TD method only requires proper temperature gradients and waiting in the presence of a temperature gradient, and any sophisticated operations are unnecessary.
Owing to the ease of execution of the method, many nanopipettes can be filled at once.
The probability of filling up of ninety-four aligned nanopipettes on the metal stage (\textcolor{blue}{Figure \ref{FIG1}}e, \href{https://pubs.acs.org/doi/suppl/10.1021/acs.analchem.9b03848/suppl_file/ac9b03848_si_001.pdf}{\textcolor{blue}{Supporting Information, Figure S2}}) was 100\% using our method.  




Time-lapsed experiments allow us to understand the filling process in the TD method.
We monitored how the filling of the nanopipettes progressed until completion (Figure \ref{FIG2}).
The nanopipette just after the injected DI water forms the meniscus at $t$ = \SI{0}{s} as 
shown in {\textcolor{blue}{Figure \ref{FIG2}}a}.
With heating the nanopipette at $t$ = \SI{5}{s}, small pools of water formed near the tip region of the nanopipette (\textcolor{blue}{Figure \ref{FIG2}}b).
With further heating of the nanopipette, water domains progressively formed, followed by the formation of small bubbles originating from the air domains, as indicated by the blue arrows in \textcolor{blue}{Figure \ref{FIG2}}b at $t$ = \SI{10}{s}.
Finally, the water filled the entire area in the nanopipette, and the air domains were completely removed at $t$ = \SI{1200}{s}. 
The formation of such small bubbles is due to the condensation of water vapor from the air/water interfaces.
In addition, the formation of a liquid bridge~\cite{zhong2016condensation,duan2012evaporation,maeda2003evaporation} would play an important role for progressively growing the water domains.
We speculate that the small bubbles move from the tip region to wide open side via the created liquid bridges, driven by the temperature gradient.
The variation in surface tension induced by the temperature gradient drives the movement of small bubbles
~\cite{miniewicz2017origin}.
The sizes of these bubbles depend on the vapor rate and pipet geometry (the height along cross-section at a certain location of a nanopipette).
Interestingly, the morphology of these bubbles varies from a cylindrical shape to a spherical or ellipsoidal one.
\begin{figure}[!t]
    \includegraphics{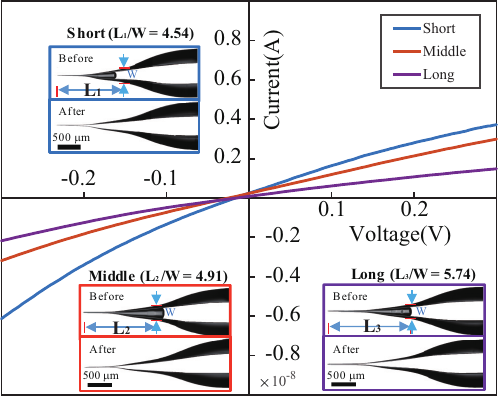}
    \caption{\label{FIG3}
    $I$-$V$ measurements for the nanopipettes with three different geometries.
    The insets show OM images before and after the application of the TD method.
    Three different nanopipettes show different $I$-$V$ shapes, indicating different conductances.
    }
\end{figure}

To show the viability of our filling method for various nanopipettes with different geometries, we performed $I$-$V$ measurements of the nanopipettes. 
We defined the aspect ratio (L/W) to quantify the difference in the pipet geometry, where L is the length along the pipet axis, measured from the pipet tip to the position at the pipet's outer diameter W = \SI{300}{\micro m} (insets of \textcolor{blue}{Figure \ref{FIG3}}; Supporting Information, \textcolor{blue}{Figure S2}).
All of the measurements were carried out with \SI{2}{M} KCl solution. 
\textcolor{blue}{Figure \ref{FIG3}} shows the $I$-$V$ curves measured for the nanopipettes with different geometries; L/W = 4.54 (short), 4.91 (middle), and 5.74 (long). 
The obtained $I$-$V$ curves showed different shapes, indicating the difference in their conductance dominantly determined by the nanopipette pore diameter and cone angle of the tip~\cite{del2014contact}.
We examined more than 300 nanopipettes with different geometries similar to those displayed in the insets of \textcolor{blue}{Figure \ref{FIG3}}.
All of them showed good electrical contacts.
This fact strongly supports the validity of the TD method regardless of the pipet geometry if the temperature gradient and waiting time are properly tuned. 

\begin{figure}[!t]
    \includegraphics{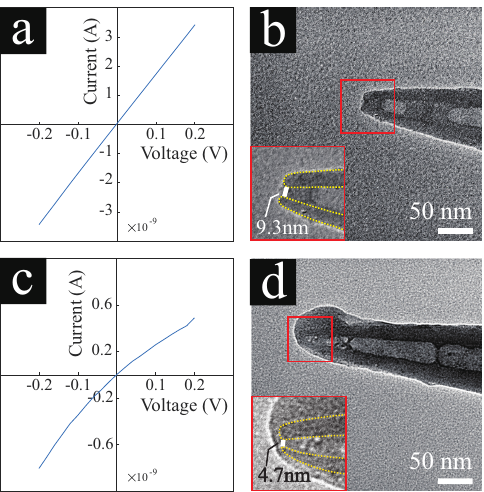}
    \caption{\label{FIG4}
    Direct evidence of the validity of the TD method for nanopipettes with pore diameters less than \SI{10}{nm}.
    The complete filling by the TD method was confirmed by $I$-$V$ measurements.
    After the confirmation of the filling, the nanopipette was imaged by TEM. (\textbf{a} and \textbf{c}) $I$-$V$ curves of two different nanopipettes used.
    (\textbf{b} and \textbf{d}) Corresponding TEM images.
    }
\end{figure}

Finally, we demonstrate the validity of the TD method for a nanopipette with a pore diameter of less than \SI{10}{nm}.
To obtain direct evidence of the filling of the nanopipette with a pore diameter less than \SI{10}{nm}, we carried out TEM measurements after measuring the $I$-$V$ curves.
\textcolor{blue}{Figure \ref{FIG4}} shows the $I$-$V$ curves and corresponding geometries of the nanopipettes captured by TEM.
The measured $I$-$V$ curves demonstrate electrical resistances of \SI{58.8}{M\Omega} (\textcolor{blue}{Figure \ref{FIG4}}a) and \SI{333.3}{M\Omega} (\textcolor{blue}{Figure \ref{FIG4}}c), indicating the formation of good electrical contacts.
A clear ion current rectification observed in \textcolor{blue}{Figure \ref{FIG4}}c indicates the formation of a very small nanopore.
This was confirmed by TEM measurements (Supporting Information, \href{https://pubs.acs.org/doi/suppl/10.1021/acs.analchem.9b03848/suppl_file/ac9b03848_si_001.pdf}{\textcolor{blue}{Figure S4}}).
The pore diameters of the nanopipettes with the low and high resistances were 9.3 (\textcolor{blue}{Figure \ref{FIG4}}b) and \SI{4.7}{nm} (\textcolor{blue}{Figure \ref{FIG4}}d), respectively.
Note that, in the TEM images, the observed particle-like structures inside the nanopipettes are difficult to remove completely.
They could be accumulated around the tip region when DI water evaporates.
The contamination from outside the nanopipettes is due to the sticky ethanol, which was utilized to fix the tip of a nanopipette onto the copper grid (see \textcolor{blue}{\hyperlink{ES}{Experimental Section}}).

Here we emphasize that the TEM measurements are quite important for accurately determining the characteristics of sub-10-nm nanopipettes.
We roughly calculated the pore diameter by the commonly used relationship between the pipet geometry and electrical properties, i.e., pore diameter, conical angle, and conductance.~\cite{del2014contact}.
The calculated pore diameters are 5.1 and \SI{1.3}{nm} for the nanopipettes displayed in \textcolor{blue}{Figure \ref{FIG4}}b and \textcolor{blue}{Figure \ref{FIG4}}d, respectively.
These values are smaller than those obtained from the TEM measurement.
This could be due to the asymmetry of the nanopipettes and/or the existence of nanobubbles, which could decrease the calculated pore diameter and are not considered in the calculation model. This research topic will be further discussed in our future work.



\section{Conclusions}

We demonstrated that the TD method is viable to fill nanopipettes with sub-10-nm pore sizes regardless of their geometrical differences.
The temperature gradient generated in the pipet removes bubbles from the pipet tip, facilitating good electrical contacts for the nanopipettes.
Unlike previous methods, our TD method is very practical and easy to introduce in nanopipette fabrication.
In addition, the TD method is a batch process, and its success rate is approximately 100\%.
This study will provide a significant contribution to various fields of nanoscience using nanopipettes.



\section*{Associated Content}

\subsection*{Supporting Information}
The Supporting Information is available free of charge on the \href{https://pubs.acs.org/}{\textcolor{blue}{ACS Publications website}} at \href{https://pubs.acs.org/doi/suppl/10.1021/acs.analchem.9b03848/suppl_file/ac9b03848_si_001.pdf}{\textcolor{blue}{DOI: 10.1021/acs.analchem.9b03848}}.\\

Pulling parameters for the fabrication of sub-10-nm pipettes, image processing of optical microscope images, setup for batch production, calculation of aspect ratios, and TEM characterization (\href{https://pubs.acs.org/doi/suppl/10.1021/acs.analchem.9b03848/suppl_file/ac9b03848_si_001.pdf}{\textcolor{blue}{PDF}})\\
\section*{Author Information}
\noindent
\textbf{Corresponding Authors}\\
Shinji Watanabe\\
E-mail: wshinji@se.kanazawa-u.ac.jp\\
Phone: (+81)76-234-4054\\

\noindent
Toshio Ando\\
E-mail: tando@staff.kanazawa-u.ac.jp\\

\noindent
\textbf{Author Contributions}\\
L.S. and K.S. contributed equally to this work.\\

\noindent
\textbf{Note}\\
The authors declare no competing financial interest.

\begin{acknowledgements}

This work was supported by a grant of JST SENTAN (JPMJSN16B4 to S.W.), Grant for Young Scientists from Hokuriku Bank (to S.W.), JSPS Grant-in-Aid for Young Scientists (B) (JP26790048 to S.W.), JSPS Grant-in-Aid for Young Scientists (A) (JP17H04818 to S.W.), JSPS Grant-in-Aid for Scientific Research on Innovative Areas (JP16H00799 to S.W.) and JSPS Grant-in-Aid for Challenging Exploratory Research (JP18K19018 to S.W.), and JSPS Grant-in-Aid for Scientific Research (S) (JP17H06121 and JP24227005 to T.A.). This work was also supported by a Kanazawa University CHOZEN project and World Premier International Research Center Initiative (WPI), MEXT, Japan.

\end{acknowledgements}



\bibliography{reference_20141021.bib}

\begin{thebibliography}{38}%
\makeatletter
\providecommand \@ifxundefined [1]{%
 \@ifx{#1\undefined}
}%
\providecommand \@ifnum [1]{%
 \ifnum #1\expandafter \@firstoftwo
 \else \expandafter \@secondoftwo
 \fi
}%
\providecommand \@ifx [1]{%
 \ifx #1\expandafter \@firstoftwo
 \else \expandafter \@secondoftwo
 \fi
}%
\providecommand \natexlab [1]{#1}%
\providecommand \enquote  [1]{``#1''}%
\providecommand \bibnamefont  [1]{#1}%
\providecommand \bibfnamefont [1]{#1}%
\providecommand \citenamefont [1]{#1}%
\providecommand \href@noop [0]{\@secondoftwo}%
\providecommand \href [0]{\begingroup \@sanitize@url \@href}%
\providecommand \@href[1]{\@@startlink{#1}\@@href}%
\providecommand \@@href[1]{\endgroup#1\@@endlink}%
\providecommand \@sanitize@url [0]{\catcode `\\12\catcode `\$12\catcode
  `\&12\catcode `\#12\catcode `\^12\catcode `\_12\catcode `\%12\relax}%
\providecommand \@@startlink[1]{}%
\providecommand \@@endlink[0]{}%
\providecommand \url  [0]{\begingroup\@sanitize@url \@url }%
\providecommand \@url [1]{\endgroup\@href {#1}{\urlprefix }}%
\providecommand \urlprefix  [0]{URL }%
\providecommand \Eprint [0]{\href }%
\providecommand \doibase [0]{http://dx.doi.org/}%
\providecommand \selectlanguage [0]{\@gobble}%
\providecommand \bibinfo  [0]{\@secondoftwo}%
\providecommand \bibfield  [0]{\@secondoftwo}%
\providecommand \translation [1]{[#1]}%
\providecommand \BibitemOpen [0]{}%
\providecommand \bibitemStop [0]{}%
\providecommand \bibitemNoStop [0]{.\EOS\space}%
\providecommand \EOS [0]{\spacefactor3000\relax}%
\providecommand \BibitemShut  [1]{\csname bibitem#1\endcsname}%
\let\auto@bib@innerbib\@empty
\bibitem [{\citenamefont {Steinbock}\ \emph {et~al.}(2013)\citenamefont
  {Steinbock}, \citenamefont {Bulushev}, \citenamefont {Krishnan},
  \citenamefont {Raillon},\ and\ \citenamefont {Radenovic}}]{steinbock2013dna}%
  \BibitemOpen
  \bibfield  {author} {\bibinfo {author} {\bibfnamefont {L.~J.}\ \bibnamefont
  {Steinbock}}, \bibinfo {author} {\bibfnamefont {R.~D.}\ \bibnamefont
  {Bulushev}}, \bibinfo {author} {\bibfnamefont {S.}~\bibnamefont {Krishnan}},
  \bibinfo {author} {\bibfnamefont {C.}~\bibnamefont {Raillon}}, \ and\
  \bibinfo {author} {\bibfnamefont {A.}~\bibnamefont {Radenovic}},\ }\href
  {https://pubs.acs.org/doi/abs/10.1021/nn405029j} {\bibfield  {journal}
  {\bibinfo  {journal} {Acs Nano}\ }\textbf {\bibinfo {volume} {7}},\ \bibinfo
  {pages} {11255} (\bibinfo {year} {2013})}\BibitemShut {NoStop}%
\bibitem [{\citenamefont {Steinbock}\ \emph {et~al.}(2010)\citenamefont
  {Steinbock}, \citenamefont {Otto}, \citenamefont {Chimerel}, \citenamefont
  {Gornall},\ and\ \citenamefont {Keyser}}]{steinbock2010detecting}%
  \BibitemOpen
  \bibfield  {author} {\bibinfo {author} {\bibfnamefont {L.~J.}\ \bibnamefont
  {Steinbock}}, \bibinfo {author} {\bibfnamefont {O.}~\bibnamefont {Otto}},
  \bibinfo {author} {\bibfnamefont {C.}~\bibnamefont {Chimerel}}, \bibinfo
  {author} {\bibfnamefont {J.}~\bibnamefont {Gornall}}, \ and\ \bibinfo
  {author} {\bibfnamefont {U.~F.}\ \bibnamefont {Keyser}},\ }\href
  {https://pubs.acs.org/doi/abs/10.1021/nl100997s} {\bibfield  {journal}
  {\bibinfo  {journal} {Nano letters}\ }\textbf {\bibinfo {volume} {10}},\
  \bibinfo {pages} {2493} (\bibinfo {year} {2010})}\BibitemShut {NoStop}%
\bibitem [{\citenamefont {Gong}\ \emph {et~al.}(2013)\citenamefont {Gong},
  \citenamefont {Patil}, \citenamefont {Ivanov}, \citenamefont {Kong},
  \citenamefont {Gibb}, \citenamefont {Dogan}, \citenamefont {deMello},\ and\
  \citenamefont {Edel}}]{gong2013label}%
  \BibitemOpen
  \bibfield  {author} {\bibinfo {author} {\bibfnamefont {X.}~\bibnamefont
  {Gong}}, \bibinfo {author} {\bibfnamefont {A.~V.}\ \bibnamefont {Patil}},
  \bibinfo {author} {\bibfnamefont {A.~P.}\ \bibnamefont {Ivanov}}, \bibinfo
  {author} {\bibfnamefont {Q.}~\bibnamefont {Kong}}, \bibinfo {author}
  {\bibfnamefont {T.}~\bibnamefont {Gibb}}, \bibinfo {author} {\bibfnamefont
  {F.}~\bibnamefont {Dogan}}, \bibinfo {author} {\bibfnamefont {A.~J.}\
  \bibnamefont {deMello}}, \ and\ \bibinfo {author} {\bibfnamefont {J.~B.}\
  \bibnamefont {Edel}},\ }\href
  {https://pubs.acs.org/doi/abs/10.1021/ac403391q} {\bibfield  {journal}
  {\bibinfo  {journal} {Analytical chemistry}\ }\textbf {\bibinfo {volume}
  {86}},\ \bibinfo {pages} {835} (\bibinfo {year} {2013})}\BibitemShut
  {NoStop}%
\bibitem [{\citenamefont {Zhang}\ \emph {et~al.}(2009)\citenamefont {Zhang},
  \citenamefont {Wood},\ and\ \citenamefont {Lee}}]{zhang2009silica}%
  \BibitemOpen
  \bibfield  {author} {\bibinfo {author} {\bibfnamefont {B.}~\bibnamefont
  {Zhang}}, \bibinfo {author} {\bibfnamefont {M.}~\bibnamefont {Wood}}, \ and\
  \bibinfo {author} {\bibfnamefont {H.}~\bibnamefont {Lee}},\ }\href
  {https://pubs.acs.org/doi/abs/10.1021/ac9009148} {\bibfield  {journal}
  {\bibinfo  {journal} {Analytical chemistry}\ }\textbf {\bibinfo {volume}
  {81}},\ \bibinfo {pages} {5541} (\bibinfo {year} {2009})}\BibitemShut
  {NoStop}%
\bibitem [{\citenamefont {Ivanov}\ \emph {et~al.}(2015)\citenamefont {Ivanov},
  \citenamefont {Actis}, \citenamefont {J{\"o}nsson}, \citenamefont
  {Klenerman}, \citenamefont {Korchev},\ and\ \citenamefont
  {Edel}}]{ivanov2015demand}%
  \BibitemOpen
  \bibfield  {author} {\bibinfo {author} {\bibfnamefont {A.~P.}\ \bibnamefont
  {Ivanov}}, \bibinfo {author} {\bibfnamefont {P.}~\bibnamefont {Actis}},
  \bibinfo {author} {\bibfnamefont {P.}~\bibnamefont {J{\"o}nsson}}, \bibinfo
  {author} {\bibfnamefont {D.}~\bibnamefont {Klenerman}}, \bibinfo {author}
  {\bibfnamefont {Y.}~\bibnamefont {Korchev}}, \ and\ \bibinfo {author}
  {\bibfnamefont {J.~B.}\ \bibnamefont {Edel}},\ }\href
  {https://pubs.acs.org/doi/abs/10.1021/acsnano.5b00911} {\bibfield  {journal}
  {\bibinfo  {journal} {ACS nano}\ }\textbf {\bibinfo {volume} {9}},\ \bibinfo
  {pages} {3587} (\bibinfo {year} {2015})}\BibitemShut {NoStop}%
\bibitem [{\citenamefont {Gibb}\ \emph {et~al.}(2014)\citenamefont {Gibb},
  \citenamefont {Ivanov}, \citenamefont {Edel},\ and\ \citenamefont
  {Albrecht}}]{gibb2014single}%
  \BibitemOpen
  \bibfield  {author} {\bibinfo {author} {\bibfnamefont {T.~R.}\ \bibnamefont
  {Gibb}}, \bibinfo {author} {\bibfnamefont {A.~P.}\ \bibnamefont {Ivanov}},
  \bibinfo {author} {\bibfnamefont {J.~B.}\ \bibnamefont {Edel}}, \ and\
  \bibinfo {author} {\bibfnamefont {T.}~\bibnamefont {Albrecht}},\ }\href
  {https://pubs.acs.org/doi/abs/10.1021/ac403921m} {\bibfield  {journal}
  {\bibinfo  {journal} {Analytical chemistry}\ }\textbf {\bibinfo {volume}
  {86}},\ \bibinfo {pages} {1864} (\bibinfo {year} {2014})}\BibitemShut
  {NoStop}%
\bibitem [{\citenamefont {Bruckbauer}\ \emph {et~al.}(2002)\citenamefont
  {Bruckbauer}, \citenamefont {Ying}, \citenamefont {Rothery}, \citenamefont
  {Zhou}, \citenamefont {Shevchuk}, \citenamefont {Abell}, \citenamefont
  {Korchev},\ and\ \citenamefont {Klenerman}}]{bruckbauer2002writing}%
  \BibitemOpen
  \bibfield  {author} {\bibinfo {author} {\bibfnamefont {A.}~\bibnamefont
  {Bruckbauer}}, \bibinfo {author} {\bibfnamefont {L.}~\bibnamefont {Ying}},
  \bibinfo {author} {\bibfnamefont {A.~M.}\ \bibnamefont {Rothery}}, \bibinfo
  {author} {\bibfnamefont {D.}~\bibnamefont {Zhou}}, \bibinfo {author}
  {\bibfnamefont {A.~I.}\ \bibnamefont {Shevchuk}}, \bibinfo {author}
  {\bibfnamefont {C.}~\bibnamefont {Abell}}, \bibinfo {author} {\bibfnamefont
  {Y.~E.}\ \bibnamefont {Korchev}}, \ and\ \bibinfo {author} {\bibfnamefont
  {D.}~\bibnamefont {Klenerman}},\ }\href
  {https://pubs.acs.org/doi/abs/10.1021/ja026816c} {\bibfield  {journal}
  {\bibinfo  {journal} {Journal of the American Chemical Society}\ }\textbf
  {\bibinfo {volume} {124}},\ \bibinfo {pages} {8810} (\bibinfo {year}
  {2002})}\BibitemShut {NoStop}%
\bibitem [{\citenamefont {Babakinejad}\ \emph {et~al.}(2013)\citenamefont
  {Babakinejad}, \citenamefont {J{\"o}nsson}, \citenamefont
  {Lo{\`p}pez~Co{\`r}doba}, \citenamefont {Actis}, \citenamefont {Novak},
  \citenamefont {Takahashi}, \citenamefont {Shevchuk}, \citenamefont {Anand},
  \citenamefont {Anand}, \citenamefont {Drews}, \citenamefont {Ferrer-Montiel},
  \citenamefont {Klenerman},\ and\ \citenamefont
  {Korchev}}]{babakinejad2013local}%
  \BibitemOpen
  \bibfield  {author} {\bibinfo {author} {\bibfnamefont {B.}~\bibnamefont
  {Babakinejad}}, \bibinfo {author} {\bibfnamefont {P.}~\bibnamefont
  {J{\"o}nsson}}, \bibinfo {author} {\bibfnamefont {A.}~\bibnamefont
  {Lo{\`p}pez~Co{\`r}doba}}, \bibinfo {author} {\bibfnamefont {P.}~\bibnamefont
  {Actis}}, \bibinfo {author} {\bibfnamefont {P.}~\bibnamefont {Novak}},
  \bibinfo {author} {\bibfnamefont {Y.}~\bibnamefont {Takahashi}}, \bibinfo
  {author} {\bibfnamefont {A.}~\bibnamefont {Shevchuk}}, \bibinfo {author}
  {\bibfnamefont {U.}~\bibnamefont {Anand}}, \bibinfo {author} {\bibfnamefont
  {P.}~\bibnamefont {Anand}}, \bibinfo {author} {\bibfnamefont
  {A.}~\bibnamefont {Drews}}, \bibinfo {author} {\bibfnamefont
  {A.}~\bibnamefont {Ferrer-Montiel}}, \bibinfo {author} {\bibfnamefont
  {D.}~\bibnamefont {Klenerman}}, \ and\ \bibinfo {author} {\bibfnamefont
  {Y.~E.}\ \bibnamefont {Korchev}},\ }\href
  {https://pubs.acs.org/doi/abs/10.1021/ac4021769} {\bibfield  {journal}
  {\bibinfo  {journal} {Analytical chemistry}\ }\textbf {\bibinfo {volume}
  {85}},\ \bibinfo {pages} {9333} (\bibinfo {year} {2013})}\BibitemShut
  {NoStop}%
\bibitem [{\citenamefont {Page}\ \emph {et~al.}(2017)\citenamefont {Page},
  \citenamefont {Kang}, \citenamefont {Armitstead}, \citenamefont {Perry},\
  and\ \citenamefont {Unwin}}]{page2017quantitative}%
  \BibitemOpen
  \bibfield  {author} {\bibinfo {author} {\bibfnamefont {A.}~\bibnamefont
  {Page}}, \bibinfo {author} {\bibfnamefont {M.}~\bibnamefont {Kang}}, \bibinfo
  {author} {\bibfnamefont {A.}~\bibnamefont {Armitstead}}, \bibinfo {author}
  {\bibfnamefont {D.}~\bibnamefont {Perry}}, \ and\ \bibinfo {author}
  {\bibfnamefont {P.~R.}\ \bibnamefont {Unwin}},\ }\href
  {https://pubs.acs.org/doi/abs/10.1021/acs.analchem.6b04629} {\bibfield
  {journal} {\bibinfo  {journal} {Analytical chemistry}\ }\textbf {\bibinfo
  {volume} {89}},\ \bibinfo {pages} {3021} (\bibinfo {year}
  {2017})}\BibitemShut {NoStop}%
\bibitem [{\citenamefont {Takahashi}\ \emph {et~al.}(2011)\citenamefont
  {Takahashi}, \citenamefont {Shevchuk}, \citenamefont {Novak}, \citenamefont
  {Zhang}, \citenamefont {Ebejer}, \citenamefont {Macpherson}, \citenamefont
  {Unwin}, \citenamefont {Pollard}, \citenamefont {Roy}, \citenamefont
  {Clifford}, \citenamefont {Shiku}, \citenamefont {Matsue}, \citenamefont
  {Klenerman},\ and\ \citenamefont {Korchev}}]{takahashi2011multifunctional}%
  \BibitemOpen
  \bibfield  {author} {\bibinfo {author} {\bibfnamefont {Y.}~\bibnamefont
  {Takahashi}}, \bibinfo {author} {\bibfnamefont {A.~I.}\ \bibnamefont
  {Shevchuk}}, \bibinfo {author} {\bibfnamefont {P.}~\bibnamefont {Novak}},
  \bibinfo {author} {\bibfnamefont {Y.}~\bibnamefont {Zhang}}, \bibinfo
  {author} {\bibfnamefont {N.}~\bibnamefont {Ebejer}}, \bibinfo {author}
  {\bibfnamefont {J.~V.}\ \bibnamefont {Macpherson}}, \bibinfo {author}
  {\bibfnamefont {P.~R.}\ \bibnamefont {Unwin}}, \bibinfo {author}
  {\bibfnamefont {A.~J.}\ \bibnamefont {Pollard}}, \bibinfo {author}
  {\bibfnamefont {D.}~\bibnamefont {Roy}}, \bibinfo {author} {\bibfnamefont
  {C.~A.}\ \bibnamefont {Clifford}}, \bibinfo {author} {\bibfnamefont
  {H.}~\bibnamefont {Shiku}}, \bibinfo {author} {\bibfnamefont
  {T.}~\bibnamefont {Matsue}}, \bibinfo {author} {\bibfnamefont
  {D.}~\bibnamefont {Klenerman}}, \ and\ \bibinfo {author} {\bibfnamefont
  {Y.~E.}\ \bibnamefont {Korchev}},\ }\href
  {https://onlinelibrary.wiley.com/doi/full/10.1002/anie.201102796} {\bibfield
  {journal} {\bibinfo  {journal} {Angewandte Chemie International Edition}\
  }\textbf {\bibinfo {volume} {50}},\ \bibinfo {pages} {9638} (\bibinfo {year}
  {2011})}\BibitemShut {NoStop}%
\bibitem [{\citenamefont {Traversi}\ \emph {et~al.}(2013)\citenamefont
  {Traversi}, \citenamefont {Raillon}, \citenamefont {Benameur}, \citenamefont
  {Liu}, \citenamefont {Khlybov}, \citenamefont {Tosun}, \citenamefont
  {Krasnozhon}, \citenamefont {Kis},\ and\ \citenamefont
  {Radenovic}}]{traversi2013detecting}%
  \BibitemOpen
  \bibfield  {author} {\bibinfo {author} {\bibfnamefont {F.}~\bibnamefont
  {Traversi}}, \bibinfo {author} {\bibfnamefont {C.}~\bibnamefont {Raillon}},
  \bibinfo {author} {\bibfnamefont {S.}~\bibnamefont {Benameur}}, \bibinfo
  {author} {\bibfnamefont {K.}~\bibnamefont {Liu}}, \bibinfo {author}
  {\bibfnamefont {S.}~\bibnamefont {Khlybov}}, \bibinfo {author} {\bibfnamefont
  {M.}~\bibnamefont {Tosun}}, \bibinfo {author} {\bibfnamefont
  {D.}~\bibnamefont {Krasnozhon}}, \bibinfo {author} {\bibfnamefont
  {A.}~\bibnamefont {Kis}}, \ and\ \bibinfo {author} {\bibfnamefont
  {A.}~\bibnamefont {Radenovic}},\ }\href
  {https://www.nature.com/articles/nnano.2013.240} {\bibfield  {journal}
  {\bibinfo  {journal} {Nature nanotechnology}\ }\textbf {\bibinfo {volume}
  {8}},\ \bibinfo {pages} {939} (\bibinfo {year} {2013})}\BibitemShut {NoStop}%
\bibitem [{\citenamefont {Venkatesan}\ and\ \citenamefont
  {Bashir}(2011)}]{venkatesan2011nanopore}%
  \BibitemOpen
  \bibfield  {author} {\bibinfo {author} {\bibfnamefont {B.~M.}\ \bibnamefont
  {Venkatesan}}\ and\ \bibinfo {author} {\bibfnamefont {R.}~\bibnamefont
  {Bashir}},\ }\href {https://www.nature.com/articles/nnano.2011.129}
  {\bibfield  {journal} {\bibinfo  {journal} {Nature nanotechnology}\ }\textbf
  {\bibinfo {volume} {6}},\ \bibinfo {pages} {615} (\bibinfo {year}
  {2011})}\BibitemShut {NoStop}%
\bibitem [{\citenamefont {Sze}\ \emph {et~al.}(2017)\citenamefont {Sze},
  \citenamefont {Ivanov}, \citenamefont {Cass},\ and\ \citenamefont
  {Edel}}]{sze2017single}%
  \BibitemOpen
  \bibfield  {author} {\bibinfo {author} {\bibfnamefont {J.~Y.}\ \bibnamefont
  {Sze}}, \bibinfo {author} {\bibfnamefont {A.~P.}\ \bibnamefont {Ivanov}},
  \bibinfo {author} {\bibfnamefont {A.~E.}\ \bibnamefont {Cass}}, \ and\
  \bibinfo {author} {\bibfnamefont {J.~B.}\ \bibnamefont {Edel}},\ }\href
  {https://www.nature.com/articles/s41467-017-01584-3} {\bibfield  {journal}
  {\bibinfo  {journal} {Nature communications}\ }\textbf {\bibinfo {volume}
  {8}},\ \bibinfo {pages} {1552} (\bibinfo {year} {2017})}\BibitemShut
  {NoStop}%
\bibitem [{\citenamefont {Wanunu}\ \emph {et~al.}(2010)\citenamefont {Wanunu},
  \citenamefont {Morrison}, \citenamefont {Rabin}, \citenamefont {Grosberg},\
  and\ \citenamefont {Meller}}]{wanunu2010electrostatic}%
  \BibitemOpen
  \bibfield  {author} {\bibinfo {author} {\bibfnamefont {M.}~\bibnamefont
  {Wanunu}}, \bibinfo {author} {\bibfnamefont {W.}~\bibnamefont {Morrison}},
  \bibinfo {author} {\bibfnamefont {Y.}~\bibnamefont {Rabin}}, \bibinfo
  {author} {\bibfnamefont {A.~Y.}\ \bibnamefont {Grosberg}}, \ and\ \bibinfo
  {author} {\bibfnamefont {A.}~\bibnamefont {Meller}},\ }\href
  {https://www.nature.com/articles/nnano.2009.379} {\bibfield  {journal}
  {\bibinfo  {journal} {Nature nanotechnology}\ }\textbf {\bibinfo {volume}
  {5}},\ \bibinfo {pages} {160} (\bibinfo {year} {2010})}\BibitemShut {NoStop}%
\bibitem [{\citenamefont {Storm}\ \emph {et~al.}(2003)\citenamefont {Storm},
  \citenamefont {Chen}, \citenamefont {Ling}, \citenamefont {Zandbergen},\ and\
  \citenamefont {Dekker}}]{storm2003fabrication}%
  \BibitemOpen
  \bibfield  {author} {\bibinfo {author} {\bibfnamefont {A.}~\bibnamefont
  {Storm}}, \bibinfo {author} {\bibfnamefont {J.}~\bibnamefont {Chen}},
  \bibinfo {author} {\bibfnamefont {X.}~\bibnamefont {Ling}}, \bibinfo {author}
  {\bibfnamefont {H.}~\bibnamefont {Zandbergen}}, \ and\ \bibinfo {author}
  {\bibfnamefont {C.}~\bibnamefont {Dekker}},\ }\href
  {https://www.nature.com/articles/nmat941} {\bibfield  {journal} {\bibinfo
  {journal} {Nature materials}\ }\textbf {\bibinfo {volume} {2}},\ \bibinfo
  {pages} {537} (\bibinfo {year} {2003})}\BibitemShut {NoStop}%
\bibitem [{\citenamefont {Li}\ \emph {et~al.}(2001)\citenamefont {Li},
  \citenamefont {Stein}, \citenamefont {McMullan}, \citenamefont {Branton},
  \citenamefont {Aziz},\ and\ \citenamefont {Golovchenko}}]{li2001ion}%
  \BibitemOpen
  \bibfield  {author} {\bibinfo {author} {\bibfnamefont {J.}~\bibnamefont
  {Li}}, \bibinfo {author} {\bibfnamefont {D.}~\bibnamefont {Stein}}, \bibinfo
  {author} {\bibfnamefont {C.}~\bibnamefont {McMullan}}, \bibinfo {author}
  {\bibfnamefont {D.}~\bibnamefont {Branton}}, \bibinfo {author} {\bibfnamefont
  {M.~J.}\ \bibnamefont {Aziz}}, \ and\ \bibinfo {author} {\bibfnamefont
  {J.~A.}\ \bibnamefont {Golovchenko}},\ }\href
  {https://www.nature.com/articles/35084037} {\bibfield  {journal} {\bibinfo
  {journal} {Nature}\ }\textbf {\bibinfo {volume} {412}},\ \bibinfo {pages}
  {166} (\bibinfo {year} {2001})}\BibitemShut {NoStop}%
\bibitem [{\citenamefont {Siwy}\ \emph {et~al.}(2003)\citenamefont {Siwy},
  \citenamefont {Dobrev}, \citenamefont {Neumann}, \citenamefont {Trautmann},\
  and\ \citenamefont {Voss}}]{siwy2003electro}%
  \BibitemOpen
  \bibfield  {author} {\bibinfo {author} {\bibfnamefont {Z.}~\bibnamefont
  {Siwy}}, \bibinfo {author} {\bibfnamefont {D.}~\bibnamefont {Dobrev}},
  \bibinfo {author} {\bibfnamefont {R.}~\bibnamefont {Neumann}}, \bibinfo
  {author} {\bibfnamefont {C.}~\bibnamefont {Trautmann}}, \ and\ \bibinfo
  {author} {\bibfnamefont {K.}~\bibnamefont {Voss}},\ }\href
  {https://link.springer.com/article/10.1007/s00339-002-1982-7} {\bibfield
  {journal} {\bibinfo  {journal} {Applied Physics A}\ }\textbf {\bibinfo
  {volume} {76}},\ \bibinfo {pages} {781} (\bibinfo {year} {2003})}\BibitemShut
  {NoStop}%
\bibitem [{\citenamefont {Rheinlaender}\ and\ \citenamefont
  {Sch{\"a}ffer}(2015)}]{rheinlaender2015lateral}%
  \BibitemOpen
  \bibfield  {author} {\bibinfo {author} {\bibfnamefont {J.}~\bibnamefont
  {Rheinlaender}}\ and\ \bibinfo {author} {\bibfnamefont {T.~E.}\ \bibnamefont
  {Sch{\"a}ffer}},\ }\href
  {https://pubs.acs.org/doi/abs/10.1021/acs.analchem.5b00900} {\bibfield
  {journal} {\bibinfo  {journal} {Analytical chemistry}\ }\textbf {\bibinfo
  {volume} {87}},\ \bibinfo {pages} {7117} (\bibinfo {year}
  {2015})}\BibitemShut {NoStop}%
\bibitem [{\citenamefont {Rheinlaender}\ and\ \citenamefont
  {Sch{\"a}ffer}(2017)}]{rheinlaender2017accurate}%
  \BibitemOpen
  \bibfield  {author} {\bibinfo {author} {\bibfnamefont {J.}~\bibnamefont
  {Rheinlaender}}\ and\ \bibinfo {author} {\bibfnamefont {T.~E.}\ \bibnamefont
  {Sch{\"a}ffer}},\ }\href
  {https://pubs.acs.org/doi/abs/10.1021/acs.analchem.7b03871} {\bibfield
  {journal} {\bibinfo  {journal} {Analytical chemistry}\ }\textbf {\bibinfo
  {volume} {89}},\ \bibinfo {pages} {11875} (\bibinfo {year}
  {2017})}\BibitemShut {NoStop}%
\bibitem [{\citenamefont {Hansma}\ \emph {et~al.}(1989)\citenamefont {Hansma},
  \citenamefont {Drake}, \citenamefont {Marti}, \citenamefont {Gould},\ and\
  \citenamefont {Prater}}]{hansma1989scanning}%
  \BibitemOpen
  \bibfield  {author} {\bibinfo {author} {\bibfnamefont {P.}~\bibnamefont
  {Hansma}}, \bibinfo {author} {\bibfnamefont {B.}~\bibnamefont {Drake}},
  \bibinfo {author} {\bibfnamefont {O.}~\bibnamefont {Marti}}, \bibinfo
  {author} {\bibfnamefont {S.}~\bibnamefont {Gould}}, \ and\ \bibinfo {author}
  {\bibfnamefont {C.}~\bibnamefont {Prater}},\ }\href
  {https://science.sciencemag.org/content/243/4891/641} {\bibfield  {journal}
  {\bibinfo  {journal} {Science}\ }\textbf {\bibinfo {volume} {243}},\ \bibinfo
  {pages} {641} (\bibinfo {year} {1989})}\BibitemShut {NoStop}%
\bibitem [{\citenamefont {Zhou}\ \emph {et~al.}(2018)\citenamefont {Zhou},
  \citenamefont {Saito}, \citenamefont {Miyamoto}, \citenamefont {Novak},
  \citenamefont {Shevchuk}, \citenamefont {Korchev}, \citenamefont {Fukuma},\
  and\ \citenamefont {Takahashi}}]{zhou2018nanoscale}%
  \BibitemOpen
  \bibfield  {author} {\bibinfo {author} {\bibfnamefont {Y.}~\bibnamefont
  {Zhou}}, \bibinfo {author} {\bibfnamefont {M.}~\bibnamefont {Saito}},
  \bibinfo {author} {\bibfnamefont {T.}~\bibnamefont {Miyamoto}}, \bibinfo
  {author} {\bibfnamefont {P.}~\bibnamefont {Novak}}, \bibinfo {author}
  {\bibfnamefont {A.~I.}\ \bibnamefont {Shevchuk}}, \bibinfo {author}
  {\bibfnamefont {Y.~E.}\ \bibnamefont {Korchev}}, \bibinfo {author}
  {\bibfnamefont {T.}~\bibnamefont {Fukuma}}, \ and\ \bibinfo {author}
  {\bibfnamefont {Y.}~\bibnamefont {Takahashi}},\ }\href
  {https://pubs.acs.org/doi/abs/10.1021/acs.analchem.7b05112} {\bibfield
  {journal} {\bibinfo  {journal} {Analytical chemistry}\ }\textbf {\bibinfo
  {volume} {90}},\ \bibinfo {pages} {2891} (\bibinfo {year}
  {2018})}\BibitemShut {NoStop}%
\bibitem [{\citenamefont {Shevchuk}\ \emph {et~al.}(2006)\citenamefont
  {Shevchuk}, \citenamefont {Frolenkov}, \citenamefont {S{\'a}nchez},
  \citenamefont {James}, \citenamefont {Freedman}, \citenamefont {Lab},
  \citenamefont {Jones}, \citenamefont {Klenerman},\ and\ \citenamefont
  {Korchev}}]{shevchuk2006imaging}%
  \BibitemOpen
  \bibfield  {author} {\bibinfo {author} {\bibfnamefont {A.~I.}\ \bibnamefont
  {Shevchuk}}, \bibinfo {author} {\bibfnamefont {G.~I.}\ \bibnamefont
  {Frolenkov}}, \bibinfo {author} {\bibfnamefont {D.}~\bibnamefont
  {S{\'a}nchez}}, \bibinfo {author} {\bibfnamefont {P.~S.}\ \bibnamefont
  {James}}, \bibinfo {author} {\bibfnamefont {N.}~\bibnamefont {Freedman}},
  \bibinfo {author} {\bibfnamefont {M.~J.}\ \bibnamefont {Lab}}, \bibinfo
  {author} {\bibfnamefont {R.}~\bibnamefont {Jones}}, \bibinfo {author}
  {\bibfnamefont {D.}~\bibnamefont {Klenerman}}, \ and\ \bibinfo {author}
  {\bibfnamefont {Y.~E.}\ \bibnamefont {Korchev}},\ }\href
  {https://onlinelibrary.wiley.com/doi/full/10.1002/anie.200503915} {\bibfield
  {journal} {\bibinfo  {journal} {Angewandte Chemie}\ }\textbf {\bibinfo
  {volume} {118}},\ \bibinfo {pages} {2270} (\bibinfo {year}
  {2006})}\BibitemShut {NoStop}%
\bibitem [{\citenamefont {Novak}\ \emph {et~al.}(2009)\citenamefont {Novak},
  \citenamefont {Li}, \citenamefont {Shevchuk}, \citenamefont {Stepanyan},
  \citenamefont {Caldwell}, \citenamefont {Hughes}, \citenamefont {Smart},
  \citenamefont {Gorelik}, \citenamefont {Ostanin}, \citenamefont {Lab},
  \citenamefont {Moss}, \citenamefont {Frolenkov}, \citenamefont {Klenerman},\
  and\ \citenamefont {Korchev}}]{novak2009nanoscale}%
  \BibitemOpen
  \bibfield  {author} {\bibinfo {author} {\bibfnamefont {P.}~\bibnamefont
  {Novak}}, \bibinfo {author} {\bibfnamefont {C.}~\bibnamefont {Li}}, \bibinfo
  {author} {\bibfnamefont {A.~I.}\ \bibnamefont {Shevchuk}}, \bibinfo {author}
  {\bibfnamefont {R.}~\bibnamefont {Stepanyan}}, \bibinfo {author}
  {\bibfnamefont {M.}~\bibnamefont {Caldwell}}, \bibinfo {author}
  {\bibfnamefont {S.}~\bibnamefont {Hughes}}, \bibinfo {author} {\bibfnamefont
  {T.~G.}\ \bibnamefont {Smart}}, \bibinfo {author} {\bibfnamefont
  {J.}~\bibnamefont {Gorelik}}, \bibinfo {author} {\bibfnamefont {V.~P.}\
  \bibnamefont {Ostanin}}, \bibinfo {author} {\bibfnamefont {M.~J.}\
  \bibnamefont {Lab}}, \bibinfo {author} {\bibfnamefont {G.~W.~J.}\
  \bibnamefont {Moss}}, \bibinfo {author} {\bibfnamefont {G.~I.}\ \bibnamefont
  {Frolenkov}}, \bibinfo {author} {\bibfnamefont {D.}~\bibnamefont
  {Klenerman}}, \ and\ \bibinfo {author} {\bibfnamefont {Y.~E.}\ \bibnamefont
  {Korchev}},\ }\href {https://www.nature.com/articles/nmeth.1306} {\bibfield
  {journal} {\bibinfo  {journal} {Nature methods}\ }\textbf {\bibinfo {volume}
  {6}},\ \bibinfo {pages} {279} (\bibinfo {year} {2009})}\BibitemShut {NoStop}%
\bibitem [{\citenamefont {Rheinlaender}\ and\ \citenamefont
  {Sch{\"a}ffer}(2013)}]{rheinlaender2013mapping}%
  \BibitemOpen
  \bibfield  {author} {\bibinfo {author} {\bibfnamefont {J.}~\bibnamefont
  {Rheinlaender}}\ and\ \bibinfo {author} {\bibfnamefont {T.~E.}\ \bibnamefont
  {Sch{\"a}ffer}},\ }\href
  {https://pubs.rsc.org/en/content/articlelanding/2013/sm/c2sm27412d/unauth#!divAbstract}
  {\bibfield  {journal} {\bibinfo  {journal} {Soft Matter}\ }\textbf {\bibinfo
  {volume} {9}},\ \bibinfo {pages} {3230} (\bibinfo {year} {2013})}\BibitemShut
  {NoStop}%
\bibitem [{\citenamefont {McKelvey}\ \emph
  {et~al.}(2014{\natexlab{a}})\citenamefont {McKelvey}, \citenamefont
  {Kinnear}, \citenamefont {Perry}, \citenamefont {Momotenko},\ and\
  \citenamefont {Unwin}}]{mckelvey2014surface}%
  \BibitemOpen
  \bibfield  {author} {\bibinfo {author} {\bibfnamefont {K.}~\bibnamefont
  {McKelvey}}, \bibinfo {author} {\bibfnamefont {S.~L.}\ \bibnamefont
  {Kinnear}}, \bibinfo {author} {\bibfnamefont {D.}~\bibnamefont {Perry}},
  \bibinfo {author} {\bibfnamefont {D.}~\bibnamefont {Momotenko}}, \ and\
  \bibinfo {author} {\bibfnamefont {P.~R.}\ \bibnamefont {Unwin}},\ }\href
  {https://pubs.acs.org/doi/abs/10.1021/ja506139u} {\bibfield  {journal}
  {\bibinfo  {journal} {Journal of the American Chemical Society}\ }\textbf
  {\bibinfo {volume} {136}},\ \bibinfo {pages} {13735} (\bibinfo {year}
  {2014}{\natexlab{a}})}\BibitemShut {NoStop}%
\bibitem [{\citenamefont {McKelvey}\ \emph
  {et~al.}(2014{\natexlab{b}})\citenamefont {McKelvey}, \citenamefont {Perry},
  \citenamefont {Byers}, \citenamefont {Colburn},\ and\ \citenamefont
  {Unwin}}]{mckelvey2014bias}%
  \BibitemOpen
  \bibfield  {author} {\bibinfo {author} {\bibfnamefont {K.}~\bibnamefont
  {McKelvey}}, \bibinfo {author} {\bibfnamefont {D.}~\bibnamefont {Perry}},
  \bibinfo {author} {\bibfnamefont {J.~C.}\ \bibnamefont {Byers}}, \bibinfo
  {author} {\bibfnamefont {A.~W.}\ \bibnamefont {Colburn}}, \ and\ \bibinfo
  {author} {\bibfnamefont {P.~R.}\ \bibnamefont {Unwin}},\ }\href
  {https://pubs.acs.org/doi/abs/10.1021/ac5003118} {\bibfield  {journal}
  {\bibinfo  {journal} {Analytical chemistry}\ }\textbf {\bibinfo {volume}
  {86}},\ \bibinfo {pages} {3639} (\bibinfo {year}
  {2014}{\natexlab{b}})}\BibitemShut {NoStop}%
\bibitem [{\citenamefont {Page}\ \emph {et~al.}(2016)\citenamefont {Page},
  \citenamefont {Perry}, \citenamefont {Young}, \citenamefont {Mitchell},
  \citenamefont {Frenguelli},\ and\ \citenamefont {Unwin}}]{page2016fast}%
  \BibitemOpen
  \bibfield  {author} {\bibinfo {author} {\bibfnamefont {A.}~\bibnamefont
  {Page}}, \bibinfo {author} {\bibfnamefont {D.}~\bibnamefont {Perry}},
  \bibinfo {author} {\bibfnamefont {P.}~\bibnamefont {Young}}, \bibinfo
  {author} {\bibfnamefont {D.}~\bibnamefont {Mitchell}}, \bibinfo {author}
  {\bibfnamefont {B.~G.}\ \bibnamefont {Frenguelli}}, \ and\ \bibinfo {author}
  {\bibfnamefont {P.~R.}\ \bibnamefont {Unwin}},\ }\href
  {https://pubs.acs.org/doi/abs/10.1021/acs.analchem.6b03744} {\bibfield
  {journal} {\bibinfo  {journal} {Analytical Chemistry}\ }\textbf {\bibinfo
  {volume} {88}},\ \bibinfo {pages} {10854} (\bibinfo {year}
  {2016})}\BibitemShut {NoStop}%
\bibitem [{\citenamefont {Perry}\ \emph {et~al.}(2015)\citenamefont {Perry},
  \citenamefont {Al~Botros}, \citenamefont {Momotenko}, \citenamefont
  {Kinnear},\ and\ \citenamefont {Unwin}}]{perry2015simultaneous}%
  \BibitemOpen
  \bibfield  {author} {\bibinfo {author} {\bibfnamefont {D.}~\bibnamefont
  {Perry}}, \bibinfo {author} {\bibfnamefont {R.}~\bibnamefont {Al~Botros}},
  \bibinfo {author} {\bibfnamefont {D.}~\bibnamefont {Momotenko}}, \bibinfo
  {author} {\bibfnamefont {S.~L.}\ \bibnamefont {Kinnear}}, \ and\ \bibinfo
  {author} {\bibfnamefont {P.~R.}\ \bibnamefont {Unwin}},\ }\href
  {https://pubs.acs.org/doi/abs/10.1021/acsnano.5b02095} {\bibfield  {journal}
  {\bibinfo  {journal} {ACS nano}\ }\textbf {\bibinfo {volume} {9}},\ \bibinfo
  {pages} {7266} (\bibinfo {year} {2015})}\BibitemShut {NoStop}%
\bibitem [{\citenamefont {Perry}\ \emph
  {et~al.}(2016{\natexlab{a}})\citenamefont {Perry}, \citenamefont
  {Paulose~Nadappuram}, \citenamefont {Momotenko}, \citenamefont {Voyias},
  \citenamefont {Page}, \citenamefont {Tripathi}, \citenamefont {Frenguelli},\
  and\ \citenamefont {Unwin}}]{perry2016surface}%
  \BibitemOpen
  \bibfield  {author} {\bibinfo {author} {\bibfnamefont {D.}~\bibnamefont
  {Perry}}, \bibinfo {author} {\bibfnamefont {B.}~\bibnamefont
  {Paulose~Nadappuram}}, \bibinfo {author} {\bibfnamefont {D.}~\bibnamefont
  {Momotenko}}, \bibinfo {author} {\bibfnamefont {P.~D.}\ \bibnamefont
  {Voyias}}, \bibinfo {author} {\bibfnamefont {A.}~\bibnamefont {Page}},
  \bibinfo {author} {\bibfnamefont {G.}~\bibnamefont {Tripathi}}, \bibinfo
  {author} {\bibfnamefont {B.~G.}\ \bibnamefont {Frenguelli}}, \ and\ \bibinfo
  {author} {\bibfnamefont {P.~R.}\ \bibnamefont {Unwin}},\ }\href
  {https://pubs.acs.org/doi/abs/10.1021/jacs.5b13153} {\bibfield  {journal}
  {\bibinfo  {journal} {Journal of the American Chemical Society}\ }\textbf
  {\bibinfo {volume} {138}},\ \bibinfo {pages} {3152} (\bibinfo {year}
  {2016}{\natexlab{a}})}\BibitemShut {NoStop}%
\bibitem [{\citenamefont {Klausen}\ \emph {et~al.}(2016)\citenamefont
  {Klausen}, \citenamefont {Fuhs},\ and\ \citenamefont
  {Dong}}]{klausen2016mapping}%
  \BibitemOpen
  \bibfield  {author} {\bibinfo {author} {\bibfnamefont {L.~H.}\ \bibnamefont
  {Klausen}}, \bibinfo {author} {\bibfnamefont {T.}~\bibnamefont {Fuhs}}, \
  and\ \bibinfo {author} {\bibfnamefont {M.}~\bibnamefont {Dong}},\ }\href
  {https://www.nature.com/articles/ncomms12447} {\bibfield  {journal} {\bibinfo
   {journal} {Nature Communications}\ }\textbf {\bibinfo {volume} {7}},\
  \bibinfo {pages} {12447} (\bibinfo {year} {2016})}\BibitemShut {NoStop}%
\bibitem [{\citenamefont {Fuhs}\ \emph {et~al.}(2018)\citenamefont {Fuhs},
  \citenamefont {Klausen}, \citenamefont {S{\o}nderskov}, \citenamefont {Han},\
  and\ \citenamefont {Dong}}]{fuhs2018direct}%
  \BibitemOpen
  \bibfield  {author} {\bibinfo {author} {\bibfnamefont {T.}~\bibnamefont
  {Fuhs}}, \bibinfo {author} {\bibfnamefont {L.~H.}\ \bibnamefont {Klausen}},
  \bibinfo {author} {\bibfnamefont {S.~M.}\ \bibnamefont {S{\o}nderskov}},
  \bibinfo {author} {\bibfnamefont {X.}~\bibnamefont {Han}}, \ and\ \bibinfo
  {author} {\bibfnamefont {M.}~\bibnamefont {Dong}},\ }\href
  {https://pubs.rsc.org/en/content/articlelanding/2018/nr/c7nr09522h/unauth#!divAbstract}
  {\bibfield  {journal} {\bibinfo  {journal} {Nanoscale}\ }\textbf {\bibinfo
  {volume} {10}},\ \bibinfo {pages} {4538} (\bibinfo {year}
  {2018})}\BibitemShut {NoStop}%
\bibitem [{\citenamefont {Salan{\c{c}}on}\ and\ \citenamefont
  {Tinland}(2018)}]{salanccon2018filling}%
  \BibitemOpen
  \bibfield  {author} {\bibinfo {author} {\bibfnamefont {E.}~\bibnamefont
  {Salan{\c{c}}on}}\ and\ \bibinfo {author} {\bibfnamefont {B.}~\bibnamefont
  {Tinland}},\ }\href
  {https://www.beilstein-journals.org/bjnano/articles/9/204} {\bibfield
  {journal} {\bibinfo  {journal} {Beilstein journal of nanotechnology}\
  }\textbf {\bibinfo {volume} {9}},\ \bibinfo {pages} {2181} (\bibinfo {year}
  {2018})}\BibitemShut {NoStop}%
\bibitem [{\citenamefont {Perry}\ \emph
  {et~al.}(2016{\natexlab{b}})\citenamefont {Perry}, \citenamefont {Momotenko},
  \citenamefont {Lazenby}, \citenamefont {Kang},\ and\ \citenamefont
  {Unwin}}]{perry2016characterization}%
  \BibitemOpen
  \bibfield  {author} {\bibinfo {author} {\bibfnamefont {D.}~\bibnamefont
  {Perry}}, \bibinfo {author} {\bibfnamefont {D.}~\bibnamefont {Momotenko}},
  \bibinfo {author} {\bibfnamefont {R.~A.}\ \bibnamefont {Lazenby}}, \bibinfo
  {author} {\bibfnamefont {M.}~\bibnamefont {Kang}}, \ and\ \bibinfo {author}
  {\bibfnamefont {P.~R.}\ \bibnamefont {Unwin}},\ }\href
  {https://pubs.acs.org/doi/abs/10.1021/acs.analchem.6b01095} {\bibfield
  {journal} {\bibinfo  {journal} {Analytical chemistry}\ }\textbf {\bibinfo
  {volume} {88}},\ \bibinfo {pages} {5523} (\bibinfo {year}
  {2016}{\natexlab{b}})}\BibitemShut {NoStop}%
\bibitem [{\citenamefont {Zhong}\ \emph {et~al.}(2016)\citenamefont {Zhong},
  \citenamefont {Zandavi}, \citenamefont {Li}, \citenamefont {Bao},
  \citenamefont {Persad}, \citenamefont {Mostowfi},\ and\ \citenamefont
  {Sinton}}]{zhong2016condensation}%
  \BibitemOpen
  \bibfield  {author} {\bibinfo {author} {\bibfnamefont {J.}~\bibnamefont
  {Zhong}}, \bibinfo {author} {\bibfnamefont {S.~H.}\ \bibnamefont {Zandavi}},
  \bibinfo {author} {\bibfnamefont {H.}~\bibnamefont {Li}}, \bibinfo {author}
  {\bibfnamefont {B.}~\bibnamefont {Bao}}, \bibinfo {author} {\bibfnamefont
  {A.~H.}\ \bibnamefont {Persad}}, \bibinfo {author} {\bibfnamefont
  {F.}~\bibnamefont {Mostowfi}}, \ and\ \bibinfo {author} {\bibfnamefont
  {D.}~\bibnamefont {Sinton}},\ }\href
  {https://pubs.acs.org/doi/abs/10.1021/acsnano.6b05666} {\bibfield  {journal}
  {\bibinfo  {journal} {ACS nano}\ }\textbf {\bibinfo {volume} {11}},\ \bibinfo
  {pages} {304} (\bibinfo {year} {2016})}\BibitemShut {NoStop}%
\bibitem [{\citenamefont {Duan}\ \emph {et~al.}(2012)\citenamefont {Duan},
  \citenamefont {Karnik}, \citenamefont {Lu},\ and\ \citenamefont
  {Majumdar}}]{duan2012evaporation}%
  \BibitemOpen
  \bibfield  {author} {\bibinfo {author} {\bibfnamefont {C.}~\bibnamefont
  {Duan}}, \bibinfo {author} {\bibfnamefont {R.}~\bibnamefont {Karnik}},
  \bibinfo {author} {\bibfnamefont {M.-C.}\ \bibnamefont {Lu}}, \ and\ \bibinfo
  {author} {\bibfnamefont {A.}~\bibnamefont {Majumdar}},\ }\href
  {https://www.pnas.org/content/109/10/3688.short} {\bibfield  {journal}
  {\bibinfo  {journal} {Proceedings of the National Academy of Sciences}\
  }\textbf {\bibinfo {volume} {109}},\ \bibinfo {pages} {3688} (\bibinfo {year}
  {2012})}\BibitemShut {NoStop}%
\bibitem [{\citenamefont {Maeda}\ \emph {et~al.}(2003)\citenamefont {Maeda},
  \citenamefont {Israelachvili},\ and\ \citenamefont
  {Kohonen}}]{maeda2003evaporation}%
  \BibitemOpen
  \bibfield  {author} {\bibinfo {author} {\bibfnamefont {N.}~\bibnamefont
  {Maeda}}, \bibinfo {author} {\bibfnamefont {J.~N.}\ \bibnamefont
  {Israelachvili}}, \ and\ \bibinfo {author} {\bibfnamefont {M.~M.}\
  \bibnamefont {Kohonen}},\ }\href
  {https://www.pnas.org/content/100/3/803.short} {\bibfield  {journal}
  {\bibinfo  {journal} {Proceedings of the National Academy of Sciences}\
  }\textbf {\bibinfo {volume} {100}},\ \bibinfo {pages} {803} (\bibinfo {year}
  {2003})}\BibitemShut {NoStop}%
\bibitem [{\citenamefont {Miniewicz}\ \emph {et~al.}(2017)\citenamefont
  {Miniewicz}, \citenamefont {Quintard}, \citenamefont {Orlikowska},\ and\
  \citenamefont {Bartkiewicz}}]{miniewicz2017origin}%
  \BibitemOpen
  \bibfield  {author} {\bibinfo {author} {\bibfnamefont {A.}~\bibnamefont
  {Miniewicz}}, \bibinfo {author} {\bibfnamefont {C.}~\bibnamefont {Quintard}},
  \bibinfo {author} {\bibfnamefont {H.}~\bibnamefont {Orlikowska}}, \ and\
  \bibinfo {author} {\bibfnamefont {S.}~\bibnamefont {Bartkiewicz}},\ }\href
  {https://pubs.rsc.org/en/content/articlehtml/2017/cp/c7cp01986f} {\bibfield
  {journal} {\bibinfo  {journal} {Physical Chemistry Chemical Physics}\
  }\textbf {\bibinfo {volume} {19}},\ \bibinfo {pages} {18695} (\bibinfo {year}
  {2017})}\BibitemShut {NoStop}%
\bibitem [{\citenamefont {Del~Linz}\ \emph {et~al.}(2014)\citenamefont
  {Del~Linz}, \citenamefont {Willman}, \citenamefont {Caldwell}, \citenamefont
  {Klenerman}, \citenamefont {Fern{\'a}ndez},\ and\ \citenamefont
  {Moss}}]{del2014contact}%
  \BibitemOpen
  \bibfield  {author} {\bibinfo {author} {\bibfnamefont {S.}~\bibnamefont
  {Del~Linz}}, \bibinfo {author} {\bibfnamefont {E.}~\bibnamefont {Willman}},
  \bibinfo {author} {\bibfnamefont {M.}~\bibnamefont {Caldwell}}, \bibinfo
  {author} {\bibfnamefont {D.}~\bibnamefont {Klenerman}}, \bibinfo {author}
  {\bibfnamefont {A.}~\bibnamefont {Fern{\'a}ndez}}, \ and\ \bibinfo {author}
  {\bibfnamefont {G.}~\bibnamefont {Moss}},\ }\href
  {https://pubs.acs.org/doi/abs/10.1021/ac402748j} {\bibfield  {journal}
  {\bibinfo  {journal} {Analytical chemistry}\ }\textbf {\bibinfo {volume}
  {86}},\ \bibinfo {pages} {2353} (\bibinfo {year} {2014})}\BibitemShut
  {NoStop}%
\end{thebibliography}%
\end{document}